# Novel Laves phase superconductor NbBe$_2$: A theoretical investigation

Md. Zahidur Rahaman[1]

*Department of Physics, Pabna University of Science and Technology,
Pabna-6600, Bangladesh*
*zahidur.physics@gmail.com*

Md. Atikur Rahman[2]

*Department of Physics, Pabna University of Science and Technology,
Pabna-6600, Bangladesh*
*atik0707phy@gmail.com*

A new Laves phase superconductor NbBe$_2$, prototype with MgCu$_2$, having maximum $T_c$ ~2.6 K has been reported very recently. Based on first-principle calculations, we systematically study the structural, elastic, mechanical, electronic, thermal and superconducting properties of the newly reported superconducting intermetallic compound NbBe$_2$. The study of Cauchy pressure and Pugh's ratio reveal the brittle manner of NbBe$_2$. This compound exhibits metallic conductivity and the Nb-Be bonds have a covalent feature in which the contribution of Nb-4d state is dominant near Fermi level. The Debye temperature of NbBe$_2$ is 664.54 K calculated using our elastic constants data. We obtain the heat capacity ($C_v$) of NbBe$_2$ at ambient condition is 20.75 cal/cell K using quasi-harmonic model. Finally, we investigate the electron-phonon coupling constant, phonon dispersion curve and density of states which indicates that the compound under study is a weakly coupled BCS superconductor.

## 1. Introduction

The Laves phase cubic MgCu$_2$ (C15) type intermetallic compound is considered to be favorable for superconductivity with widely varying transition temperatures, ranging from 0.07 K to above 10 K [1]. Due to these large differences in superconducting critical temperature in the same crystal classes, the details study of these compounds is quite interesting. To understand the nature of superconductivity properly in different superconductors still remain a challenge of modern condensed matter physics due to the complication of the structure, magnetism and electronic nematicity. Since the structure and the chemical composition of Laves phase compounds are quite simple, hence the details study of these compounds give a deep insight for further understanding the fundamental nature of superconductivity. In 1974 O. Rapp investigated the superconductivity in different Laves phase compounds and reported the superconducting temperature of ZrW$_2$, HfW$_2$ and ZrMo$_2$ Lave phases are 0.12 K [1]. New niobium beryllide Nb$_3$Be with the A15 type structure has been synthesized by Tuleushev et al. using thermal treatment of the amorphous film system [2]. He reported the superconductivity in Nb$_3$Be is 10 K which is quite large. The chemical composition of NbBe$_2$ is alike with Nb$_3$Be and hence it is quite interesting to investigate the superconductivity in NbBe$_2$ system.

------------------------------------------------------------

[2]Corresponding Author: atik0707phy@gmail.com

The NbBe$_2$ Laves phase compound prototype with MgCu$_2$ was synthesized by Donald E. Sands in 1959 [3]. He reported that NbBe$_2$ has a face centered cubic cell with lattice parameter a = 6.535 Å containing eight formula unit per unit cell. Later in 2015 H. Hosono et al further synthesized the NbBe$_2$ intermetallic and reported the superconductivity of this compound with the transition temperature $T_c$ ~2.6 K [4].

However no experimental and theoretical information are available yet on the physical properties of this superconductor. We therefore decided to investigate the details physical properties of this compound theoretically. In this paper we have calculated different physical properties of NbBe$_2$ intermetallic including structural, elastic, mechanical, electronic, thermal and superconducting properties by using the plane-wave pseudopotential density functional theory method (DFT) with the aim of having a profound comprehension about these properties and then the data are analyzed systematically. The remaining parts of this paper are organized as follows. The theoretical methods are discussed in section 2, the investigated results and the related discussions are presented in section 3 and finally, a summary of this present work is shown in section 4.

## 2. Computational details

We have carried out our investigation by using the density functional theory based CASTEP computer program together with the generalized gradient approximation (GGA) with the PBESOL exchange-correlation function [5-9]. The pseudo atomic calculation is performed for Nb-4s$^2$ 4p$^6$ 4d$^4$ 5s$^1$ and Be-2s$^2$. The k-point sampling of the Brillouin zone was constructed using Monkhorst-Pack scheme [10] with 4×4×4 grids in primitive cells of NbBe$_2$. The value of the energy cut-off was set to 320 eV and the electromagnetic wave functions were expanded in a plane wave basis set with this energy. The equilibrium crystal structures were obtained via geometry optimization in the Broyden-Fletcher-Goldfarb-Shanno (BFGS) minimization scheme [11]. For obtaining the optimized structure of NbBe$_2$ criteria of convergence were set to $1.0 \times 10^{-5}$ eV/atom for energy, $1 \times 10^{-3}$ Å for ionic displacement, 0.03 eV/ Å for force and 0.05 GPa for stress.

The elastic stiffness constants of cubic NbBe$_2$ were obtained via the stress-strain method [12] at the optimized structure under the condition of each pressure. In this case the criteria of convergence were set to $2.0 \times 10^{-6}$ eV/atom for energy, $2.0 \times 10^{-4}$ Å for maximum ionic displacement and 0.006 eV/ Å for maximum ionic force. The value of the maximum strain amplitude was set to be 0.003 for this investigation. The Debye temperature of this compound was calculated using our elastic constant data.

## 3. Results and discussion

### 3.1. Structural properties

NbBe$_2$ typically possesses a face centered cubic crystal structure with the space group Fd-3m (227) and has an equilibrium lattice parameter 6.535 Å [3]. It belongs to MgCu$_2$ type structure where the atomic positions of Nb and Be atom in the unit cell are (0, 0, 0) and (0.625, 0.625, 0.625) respectively. A unit cell of NbBe$_2$ typically contains eight formula units in which each Nb atom is surrounded by 12 Be atoms at 2.71 Å and 4 Nb atoms at 2.83 Å [3]. We optimize the lattice parameter and atomic positions of this intermetallic as a function of normal stress by minimizing the total energy as shown in Fig.1. The investigated structural parameters are listed in Table 1. The calculated lattice constant of NbBe$_2$ is 6.484 Å which shows actually very minor deviation (0.78%) from the experimental value and evidently bears the reliability of our present DFT based investigation.

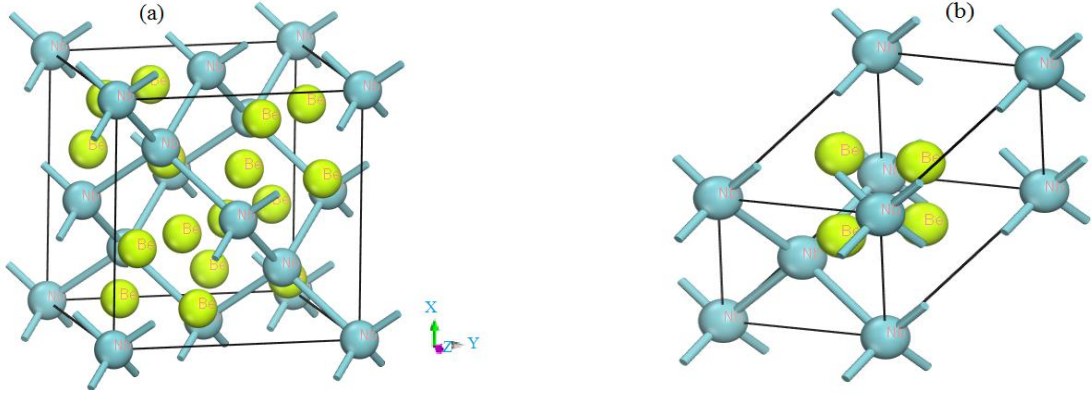

**Fig. 1.** The crystal structures of NbBe$_2$ (a) conventional cubic cell and (b) primitive cell.

**Table 1.** The calculated equilibrium Lattice constant "$a_0$", unit cell volume "$V_0$" and bulk modulus "$B_0$" of NbBe$_2$ superconductor.

| Properties | Expt.[6] | Other Calculation | Present Calculation | Deviation from Expt. (%) |
|---|---|---|---|---|
| $a_0$ (Å) | 6.535 | - | 6.484 | 0.78 |
| $V_0$ (Å$^3$) | - | - | 272.26 | - |
| $B_0$ (GPa) | - | - | 254.56 | - |

## 3.2. Elastic Properties

Elastic constants are very important materials parameters to understand the various fundamental solid-state phenomena such as Debye temperature, chemical bonds, and the mechanical stability of materials. The details study of elastic constants provides a link between the mechanical properties and dynamic information concerning the nature of the forces operating in solids, especially for the stability and stiffness of materials [13].The details study of elastic constants provides lucid idea about various materials properties such as stiffness, brittleness, stability, ductility, and anisotropy of material. In this section we have done a details study about the elastic constants and mechanical properties of NbBe$_2$ Laves phase. The elastic constants were obtained from a linear fit of the calculated stress-strain function according to Hook's law [14]. The intermetallic NbBe$_2$ has three independent elastic constants $C_{11}$, $C_{12}$ and $C_{44}$ since this compound belongs to cubic crystal structure. Using the value of calculated $C_{ij}$, the most important mechanical properties such as the bulk modulus $B$, shear modulus $G$, Young's modulus $E$ and Poisson's ratio $v$ of NbBe$_2$ are determined by using the Voigt-Reuss-Hill (VRH) averaging scheme [15]. For the cubic system, the Voigt and Reuss bounds of $B$ and $G$ can be expressed as follows [16]:

$$B_v = B_R = \frac{(C_{11} + 2C_{12})}{3} \qquad (1)$$

$$G_v = \frac{(C_{11} - C_{12} + 3C_{44})}{5} \qquad (2)$$

$$G_R = \frac{5C_{44}(C_{11} - C_{12})}{[4C_{44} + 3(C_{11} - C_{12})]} \tag{3}$$

The Hill took an arithmetic mean value of *B* and *G* can be expressed as follows,

$$B = \frac{1}{2}(B_R + B_v) \tag{4}$$

$$G = \frac{1}{2}(G_v + G_R) \tag{5}$$

Young's modulus (*E*), and Poisson's ratio (*v*) can be calculated by using following relations,

$$E = \frac{9GB}{3B + G} \tag{6}$$

$$v = \frac{3B - 2G}{2(3B + G)} \tag{7}$$

The Zener anisotropy factor *A* is calculated by using the following formula-

$$A = \frac{2C_{44}}{(C_{11} - C_{12})} \tag{8}$$

The calculated values of $C_{ij}$, *B*, *G*, *E*, *v*, *A* and *B/G* at ambient pressure are listed in Table 2. We are not being able to compare our calculated data with others because no other experimental and theoretical data are available in literature yet.

**Table 2.** The calculated elastic constants $C_{ij}$ (GPa), Cauchy pressure ($C_{12} - C_{44}$), bulk modulus *B* (GPa), shear modulus *G* (GPa), Young's modulus *E* (GPa), *B/G* values, Poisson's ratio *v* and anisotropy factor *A* of NbBe$_2$ intermetallic.

| $C_{11}$ | $C_{12}$ | $C_{44}$ | $C_{12}$-$C_{44}$ | B | G | E | B/G | v | A |
|---|---|---|---|---|---|---|---|---|---|
| 309.50 | 83.22 | 109.46 | -26.24 | 158.65 | 110.93 | 269.88 | 1.43 | 0.21 | 0.97 |

The well known Born stability criteria [17] should be satisfied for mechanically stable crystals. For cubic crystal structure the criteria are as follows-

$$C_{11} > 0, C_{44} > 0, C_{11} - C_{12} > 0 \text{ and } C_{11} + 2C_{12} > 0$$

From table 2 we see that the above stability criteria are satisfied by our calculated elastic constants. Hence NbBe$_2$ is mechanically stable. The Cauchy pressure is defined as $C_{12} - C_{44}$ and is used to describe the angular character of atomic bonding [18]. For a nonmetallic compound the value of the Cauchy pressure is negative and is positive for metallic compound [19]. From table 2, we see that the compound under study has a negative value of Cauchy pressure which indicates the nonmetallic behavior of NbBe$_2$. This result shows direct contradiction with the result having from the analysis of electronic band structure of this compound discussed in section 3.3. The reason of this contradiction is currently unknown and hopes to investigate in future. Furthermore the negative value of Cauchy's pressure indicates the brittle nature of NbBe$_2$ [20]. Pugh's ratio is another useful index to explain the ductility and brittleness of a material and is defined as *B/G* [21]. The value of *B/G* < 1.75 indicates the brittle nature of a compound otherwise the compound will be ductile. Our calculated Pugh's ratio of $NbBe_2$ intermetallic is 1.43 which indicates the brittle manner of $NbBe_2$. This result is similar with the result obtained from Cauchy pressure indicating the reliability of our work.

From Table 2, it is evident that the value of bulk modulus is larger than the value of shear modulus, which indicates that the shear modulus is the prominent parameter associating with the stability of $NbBe_2$ superconductor [22]. The value of the bulk modulus $B$ calculated using our elastic constant data match well with $B_0$, listed in Table 1, which was obtained through the fit to a Birch-Murnaghan EOS. The larger value of bulk modulus indicates the stronger capacity of resist deformation. On the other hand, the shear moduli and Young's modulus are the measure of resist reversible deformation by shear stress and stiffness of the solid materials respectively [23]. The calculated values of $B$, $G$ and $E$ of $NbBe_2$ are 158.65, 110.93 and 269.88 GPa respectively listed in table 2. The Poisson's ratio is another useful parameter to understand the nature of bonding force in a material [24]. The smaller value of $v$ ($v = 0.1$) indicates the covalent materials whereas for ionic crystal $v = 0.25$. The value between 0.25 and 0.5 denotes the force exists in a compound is central [25]. From Table 2, we see that the value of $v$ is 0.21 at ambient condition exposing the domination of covalent nature in $NbBe_2$ intermetallic. The Zener anisotropy factor $A$ is the measure of the degree of anisotropy in solid [26]. The value of $A$ for a completely isotropic material is 1. On the other hand a value smaller or larger than 1 indicates the degree of elastic anisotropy. The calculated value of $A$ for $NbBe_2$ is 0.97, indicating that $NbBe_2$ is an elastically anisotropic material.

### 3.3. Electronic properties

The electronic properties of $NbBe_2$ intermetallic compound is studied by calculating the electronic band structure, partial density of states (PDOS) and total density of states (TDOS) as shown in Fig. 2 and Fig. 3 respectively. From Fig. 2 we see that the conduction and valence bands are overlapped with each other which indicating that the compound under study is metallic in nature. The metallic nature of $NbBe_2$ indicates that this compound might be a superconductor.

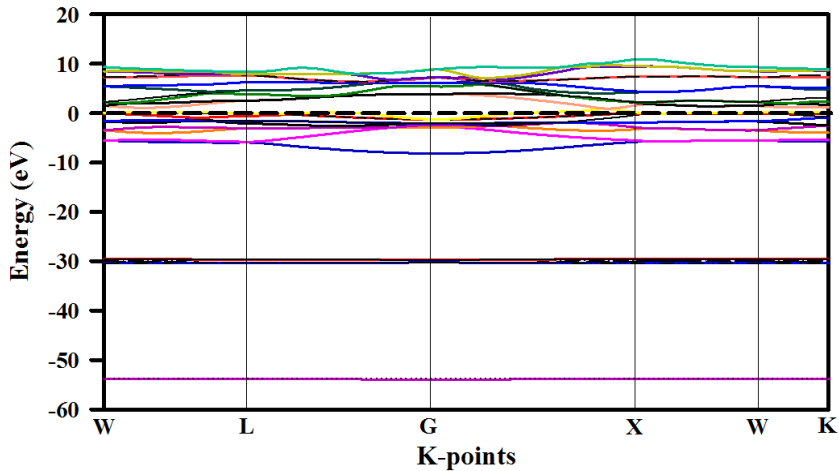

**Fig. 2.** Electronic band structure of $NbBe_2$ superconductor along high symmetry direction in the Brillouin zones.

The predominant feature of hybridization for Be-2s and Nb-4d orbital is observed from 8 eV to Fermi level as shown in Fig. 3. However near Fermi level the contribution of Nb-4d state is dominant. In the PDOS diagram we observe two pseudogap located above the Fermi level at around 0.73 eV and 6.34 eV respectively. In conduction band the contribution of Nb-4d state is dominant up to 6.34 eV and the contribution of Be-2s state is dominant in the later part. It is also evident from Fig. 3 that the PDOS of Be-2s states coincide with those of Nb-4d states below the Fermi level, which implies that the strong interaction between Nb and Be atom is highly covalent. The calculated density of states of $NbBe_2$ at Fermi level is 3.86 electrons/eV.

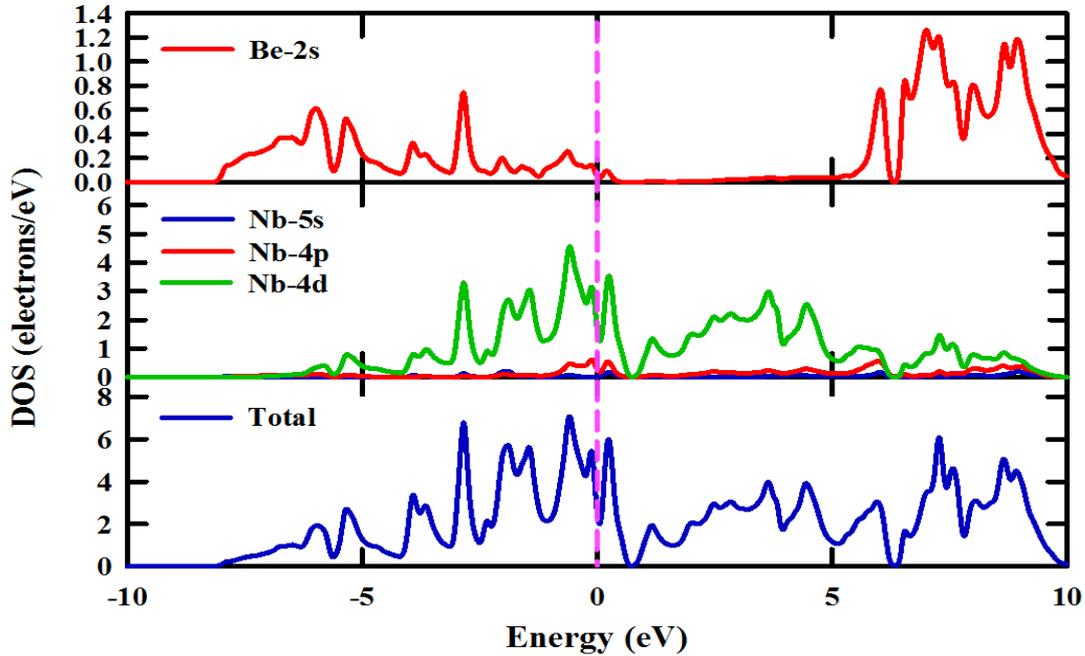

**Fig. 3.** The partial and total density of states of $NbBe_2$

In order to further explore the bonding and analyze the covalent or ionic character of $NbBe_2$ superconductor, Mulliken overlap population [27] is investigated and listed in Table 3. It is a great quantitative criterion for further understanding the bonding property of materials. A value of zero of the bond population indicates a perfectly ionic bond whereas the values greater than zero indicate the increasing levels of covalency [28]. A value less than zero also indicate the ionic bond.

**Table 3.** Mulliken atomic populations of $NbBe_2$.

| Species | s | p | d | Total | Charge | Bond | Population | Lengths (Å) |
|---|---|---|---|---|---|---|---|---|
| Be | 0.44 | 1.88 | 0.00 | 2.32 | -0.32 | Be-Be | 0.45 | 2.292 (2.310)[a] |
| Nb | 2.07 | 6.26 | 4.04 | 12.36 | 0.64 | Be-Nb | 0.33 | 2.688 (2.709)[a] |
| | | | | | | Nb-Nb | 0.33 | 2.807 (2.830)[a] |

[a]Ref. 3

From Table 2 we see that charge transfers from Nb to Be atom. The bond population of $NbBe_2$ is positive and greater than zero indicating the covalent nature of this compound which is agreeable with the result having from the analysis of DOS. Overall, the study of DOS and Mulliken atomic population reveal that the covalent bonds are formed between Nb and Be atom, and $NbBe_2$ superconductor is classified as a covalent crystal. The interatomic distances in $NbBe_2$ are also evaluated and listed in Table 1. Our calculated bond lengths are very near to the experimental values which indicate the reliability of the present calculation.

*3.4. Thermodynamic Properties*

It is very important to investigate the thermodynamic properties of materials which lead to profound understanding about many crucial solid-state phenomena directly. In this section, we have studied the temperature dependence thermodynamic properties of NbBe$_2$ superconductor up to 1000 K by using the Quasi-harmonic approximation. The investigated phonon dispersion relation and phonon density of states are shown in Fig. 4. The change of the enthalpy, entropy, free energy and heat capacity with temperature are shown in Fig. 5.

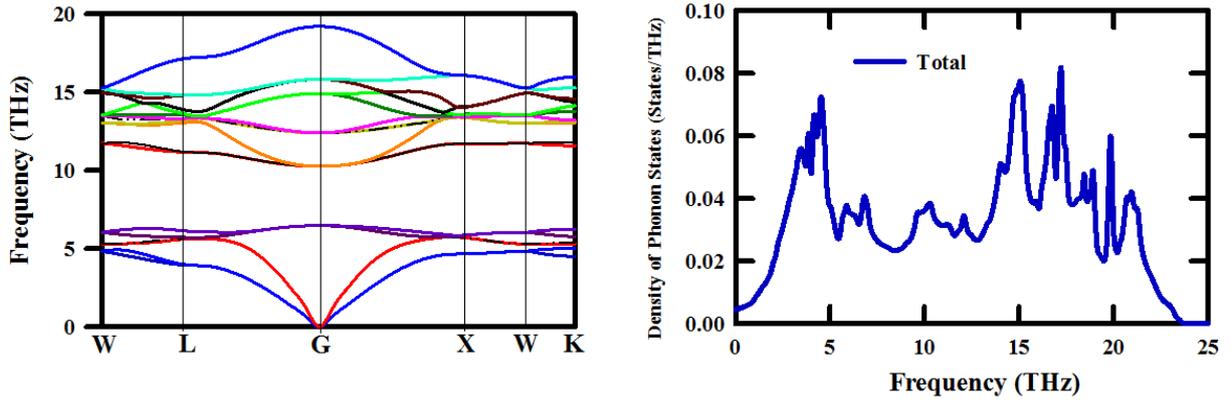

**Fig. 4.** The calculated phonon dispersion relations and total phonon density of states of NbBe$_2$.

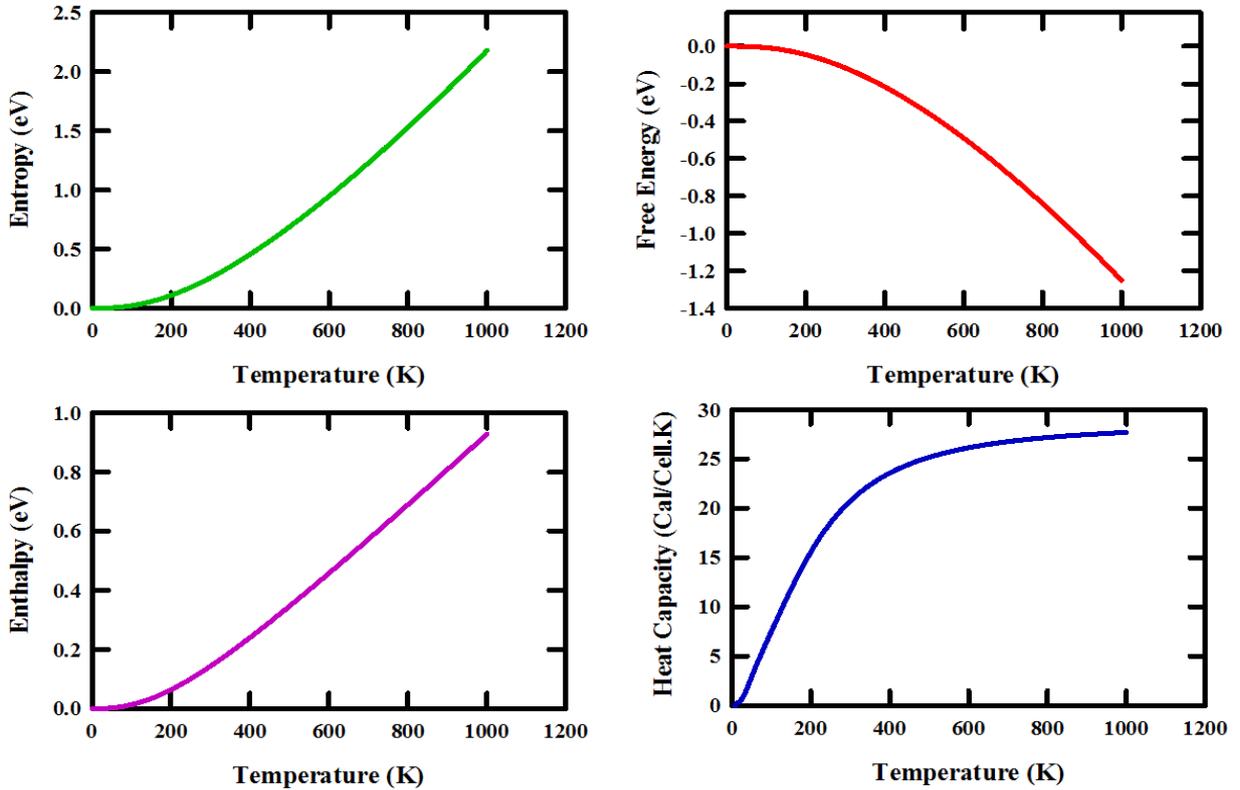

**Fig. 5.** Temperature dependence of the entropy, free energy, enthalpy and heat capacity at constant volume for NbBe$_2$ superconductor.

It is evident from Fig. 5 that the entropy, enthalpy and heat capacity increase with increasing temperature while the free energy decreases with temperature. We observe that the heat capacity of NbBe$_2$ intermetallic increases more rapidly than the entropy and enthalpy, which indicates that the heat capacity of NbBe$_2$ superconductor is more temperature sensitive than entropy and enthalpy. From Fig. 5 the calculated values of entropy, enthalpy and free energy at room temperature are 0.26 eV, 0.14 eV and -0.10 eV respectively. Generally lattice or phonon and electron both contribute to the heat capacity. The domination of the phonon contribution is strong at low temperature while at high temperature the electron contribution is more important. It is clear from Fig. 5 that the heat capacity at constant volume C$_v$ is proportional to T$^3$ at low temperature range (T < 300 K), and increases gradually with temperature and finally reaches a constant value 27.65 cal/cell K which is known as Dulonge Petit limit. At ambient condition the value of C$_v$ is 20.75 cal/cell K for NbBe$_2$ superconductor.

The Debye temperature ($\Theta_D$) is a very crucial thermodynamic parameter for the measurement of crystals physical properties such as specific heat, thermal expansion, melting point etc. Hence it is necessary to determine the Debye temperature of NbBe$_2$ superconductor. In Debye theory, the Debye temperature is the temperature of a crystal's highest normal mode of vibration. Hence Debye temperature $\Theta_D$ is the highest temperature that can be achieved due to a single normal lattice vibration. The Debye temperature can be calculated by using the elastic constants data, which is same as that determined from specific heat measurement of solid at low temperature. The Debye temperature $\Theta_D$ is obtained from the average sound velocities V$_m$ and unit-cell by using the following equation [29]:

$$\theta_D = \frac{h}{k_B}\left(\frac{3N}{4\pi V}\right)^{\frac{1}{3}} \times v_m \tag{9}$$

Where, h is defined as the Planck constant, N is the number of atoms in unit cell and V is the volume of the unit cell. In the above equation V$_m$ is the average sound velocity, which is given by-

$$v_m = \left[\frac{1}{3}\left(\frac{2}{v_t^3} + \frac{1}{v_l^3}\right)\right]^{-\frac{1}{3}} \tag{10}$$

Where, $V_l$ and $V_t$ are longitudinal and transverse elastic wave velocities respectively, which can be expressed as follows:

$$v_t = \left(\frac{G}{\rho}\right)^{\frac{1}{2}} \tag{11}$$

And

$$v_l = \left(\frac{3B + 4G}{3\rho}\right)^{\frac{1}{2}} \tag{12}$$

The calculated transverse, longitudinal and average sound velocities of NbBe$_2$ superconductor using the above equations are tabulated in Table 4. By using these values in Eq. (9) the calculated Debye temperature of NbBe$_2$ superconductor is 664.54 K. As far as we know, there are no experimental and other theoretical data available in literature for comparison our present calculated data. Hence we are unable to evaluate the magnitude of errors between theory and experiment. Therefore, we consider the present results as a prediction which awaits an experimental confirmation in future.

**Table 4.** The calculated density $\rho$ (gm/cm$^3$), transverse ($V_t$), longitudinal ($V_l$), and average sound velocity $V_m$ (m/s) and Debye temperature $\Theta_D$ (K) of NbBe$_2$ compound.

|  | $\rho$ | $V_t$ | $V_l$ | $V_m$ | $\Theta_D$ |
|---|---|---|---|---|---|
| **This work** | 5.40 | 4532.12 | 7534.23 | 5012.37 | 664.54 |
| **Expt.** | 5.28[a] | - | - | - | - |

[a]Ref. 3

*3.5. Electron-phonon coupling constant and superconductivity*

For evaluating the superconducting transition temperature $T_c$ of NbBe$_2$ intermetallic we need to calculate the electron-phonon coupling constant λ accurately. In general QUANTUM-ESPRESSO program is used to compute the value of λ directly [30]. But the problem arises in this process is that a double-delta function integration over a dense net of electron and phonon λ and q vectors is required to compute λ accurately even in case of simple compound [31]. Hence it requires significant computational resources. As an alternative method, we can compute the value of λ by using the following equation, [32]

$$1 + \lambda = \frac{3\gamma}{[2\pi^2 k_b N(E_F)]} \quad (13)$$

Where, N ($E_F$) is the density of state (DOS) at Fermi level and γ is the electronic specific-heat coefficient. We can get the value of DOS at Fermi level from our present DOS analysis of NbBe$_2$ but there is no experimental value of γ is available in literature for NbBe$_2$ and the theoretically evaluated value will not provide good estimation as this is known to produce lower value of γ [32]. Since it is not possible for us to evaluate the value of λ accurately hence we can't accurately calculate the value of $T_c$ for the compound under consideration. Considering the above discussion we compute the value of λ indirectly using McMillan's equation [33] as follows:

$$\lambda = \frac{1.04 + \mu^* \ln\left(\frac{\theta_D}{1.45 T_c}\right)}{(1 - 0.62\mu^*) \ln\left(\frac{\theta_D}{1.45 T_c}\right) - 1.04} \quad (14)$$

Where, $\mu^*$ is the effective screened Coulomb repulsion constant, whose value lies between 0.10 and 0.16. In our present calculation we take 0.13 as the value of $\mu^*$. Using our calculated Debye temperature $\theta_D$ = 664.54 K, experimentally measured $T_c$ = 2.6 K and $\mu^*$ = 0.13 in the above equation we have the value of electron-phonon coupling constant λ = 0.46 for NbBe$_2$. Since no experimental and theoretical work is available in literature yet hence we are not being able to compare our calculated result to ensure the high accuracy. However, this value of λ implies that NbBe$_2$ is typically a weakly coupled Bardeen-Copper-Schrieffer (BCS) superconductor.

## 4. Conclusions

In summary, in this work we have extensively investigated the details physical and superconducting properties of recently reported Laves phase superconductor NbBe$_2$ by using *ab initio* evolutionary simulation method. The study on elastic properties ensures that NbBe$_2$ is mechanically stable compound and brittle in manner. The electronic band structure shows that the compound under study is a metal. Analyzing the PDOS and Mulliken atomic populations we conclude that NbBe$_2$ can be classified as a covalent crystal and the calculated value of Poisson's ratio strengthen this prediction. We also determine the Debye temperature and heat capacity of this intermetallic through the investigation of details thermodynamic properties. Furthermore the electron-phonon coupling constant λ has been calculated and according to which we predict that NbBe$_2$ is a weakly coupled BCS superconductor. The present study has a great implication for future investigation on the physical and superconducting properties of others laves phase compounds.

## References

[1] Rapp, Ö., J. Invarsson, and T. Claeson. "Search for superconductivity in Laves phase compounds." *Physics Letters A* 50.3 (1974): 159-160.

[2] Tuleushev, A. Zh, V. N. Volodin, and Yu Zh Tuleushev. "Novel superconducting niobium beryllide Nb3Be with A15 structure." *Journal of Experimental and Theoretical Physics Letters* 78.7 (2003): 440-442.

[3] Sands, DONALD E., A. Zalkin, and O. H. Krikorian. "The crystal structure of NbBe2 and NbBe3." *Acta Crystallographica* 12.6 (1959): 461-464.

[4] Hosono, Hideo, et al. "Exploration of new superconductors and functional materials, and fabrication of superconducting tapes and wires of iron pnictides." *Science and Technology of Advanced Materials* 16.3 (2015): 033503.

[5] S.J. Clark, M.D. Segall, C.J. Pickard, P.J. Hasnip, M.J. Probert, K. Refson, M.C. Payne, Z.Kristallogr. 220 (2005) 567–570.

[6] Materials Studio CASTEP manual_Accelrys, 2010. pp. 261–262. <http:// www.tcm.phy.cam.ac.uk/castep/documentation/WebHelp/CASTEP.html>.

[7] P. Hohenberg, W. Kohn, Phys. Rev. 136 (1964) B864–B871.

[8] J.P. Perdew, A. Ruzsinszky, G.I. Csonka, O.A. Vydrov, G.E. Scuseria, L.A. Constantin, X. Zhou, K. Burke, Phys. Rev. Lett. 100 (2008) 136406–136409.

[9] J.P. Perdew, A. Ruzsinszky, G.I. Csonka, O.A. Vydrov, G.E. Scuseria, L.A. Constantin, X. Zhou, K. Burke, Phys. Rev. Lett. 100 (2008) 136406.

[10] H. J. Monkhorst and J. D. Pack, Phys. Rev. B 13, 5188 (1976).

[11] B. G. Pfrommer, M. Cote, S. G. Louie, and M. L. Cohen, J. Comput. Phys. 131, 233 (1997).

[12] J. Kang, E.C. Lee, K.J. Chang, Phys. Rev. B 68 (2003) 054106.

[13] Wang JY, Zhou YC. Physical Review B 2004;69:214111-9.

[14] J.F. Nye, Proprie´te´s Physiques des Mate´riaux, Dunod, Paris, 1961.

[15] Hill R. The elastic behaviour of a crystalline aggregate. Proceedings of the Physical Society A 1952;65:349-54.

[16] Wu ZJ, Zhao EJ, Xiang HP, Hao XF, Liu XJ, Meng J. Physical Review B 2007;76:054115-29.

[17] M. Born, in On the Stability of Crystal Lattices. I (Cambridge University Press, 1940), p. 160.

[18] Pettiifor DG. Journal of Materials Science and Technology 1992;8:345-9.

[19] Yong Liu, Wen-Cheng Hu, De-jiang Li, Xiao-Qin Zeng, Chun-Shui Xu, Xiang-Jie Yang, Intermetallics, 31 (2012) 257-263.

[20]  D. Pettifor, Mater. Sci. Technol. 8 (1992) 345–349.
[21]  S.F. Pugh, Philos. Mag. 45 (1954) 823–843.
[22]  Q.J. Liu, Z.T. Liu, L.P. Feng, H. Tian, Comput. Mater. Sci. 50 (2011) 2822.
[23]  Rahman, Md, and Md Rahaman. "The structural, elastic, electronic and optical properties of MgCu under pressure: A first-principles study." *arXiv preprint arXiv:1510.02020* (2015).
[24]  Y. Cao, J.C. Zhu, Y. Liu, Z.S. Nong, Z.H. Lai, Comput. Mater. Sci. 69 (2013) 40.
[25]  B. G. Pfrommer, M. C^ot_e, S. G. Louie, and M. L. Cohen, J. Comput. Phys. 131, 233 (1997).
[26]  C. Zener, Elasticity and Anelasticity of Metals, University of Chicago Press, Chicago, 1948.
[27]  R.S. Mulliken, J. Chem. Phys. 23 (1955) p.1833.
[28]  Segall, M. D.; Shah, R.; Pickard, C. J.; Payne, M. C. *Phys. Rev. B*, 54, 16317-16320 (1996).
[29]  S. Aydin, M. Simsek, Phys. Rev. B: Condens. Matter 80 (2009) 134107.
[30]  P. Giannozzi, S. Baroni, N. Bonini, M. Calandra, R. Car, C. Cavazzoni, D. Ceresoli, G.L. Chiarotti, M. Cococcioni, I. Dabo, A. Dal Corso, S. Fabris, G. Fratesi, S. de Gironcoli, R. Gebauer, U. Gerstmann, C. Gougoussis, A. Kokalj, M. Lazzeri, L. Martin-Samos, N. Marzari, F. Mauri, R. Mazzarello, S. Paolini, A. Pasquarello, L. Paulatto, C. Sbraccia, S. Scandolo, G. Sclauzero, A.P. Seitsonen, A. Smogunov, P. Umari, R.M. Wentzcovitch, J. Phys.:Condens. Matter 21 (2009) 395502, http://dx.doi.org/10.1088/0953-8984/21/39/395502.
[31]  Ali, M. S., M. Aftabuzzaman, M. Roknuzzaman, M. A. Rayhan, F. Parvin, M. M. Ali, M. H. K. Rubel, and A. K. M. A. Islam. "New superconductor (Na 0.25 K 0.45) Ba 3 Bi 4 O 12: A first-principles study." *Physica C: Superconductivity and its Applications* 506 (2014): 53-58.
[32]  S.Y. Savrasov, D.Y. Savrasov, Phys. Rev. B. 54 (23) (1996) 16487–16501.
[33]  W.L. McMillan, Phys. Rev. 167 (1968) 331–344.